\documentclass[%
twocolumn,
amsmath,
amssymb,
aps,
prl,
nofootinbib,
superscriptaddress,
floatfix,
longbibliography]{revtex4-1}

\usepackage[pdftex]{graphicx}
\usepackage{hyperref}

\usepackage{bm}
\usepackage{braket}
\usepackage{mathtools}
\usepackage{xspace}
\usepackage{array}
\usepackage{multirow}
\usepackage{comment}

\hypersetup{colorlinks = true, urlcolor = blue, linkcolor = blue, citecolor = blue}

\makeatletter

\let\MYcaption\@makecaption
\makeatother

\usepackage{subcaption}
\makeatletter
\let\@makecaption\MYcaption
\makeatother

\usepackage{color}
\usepackage[normalem]{ulem} 

\newcommand{\bk}{\bm{k}}

\begin{document}

\title{Even-odd parity transition in strongly correlated locally noncentrosymmetric superconductors : An application to CeRh$_2$As$_2$}

\author{Kosuke Nogaki}
\email[]{nogaki.kosuke.83v@st.kyoto-u.ac.jp}
\affiliation{%
  Department of Physics, Kyoto University, Kyoto 606-8502, Japan
}%

\author{Youichi Yanase}
\affiliation{%
  Department of Physics, Kyoto University, Kyoto 606-8502, Japan
}%

\date{\today}

\begin{abstract}
The discovery of the multiple $H$-$T$ phase diagram of CeRh$_2$As$_2$ offers a new route to designing topological superconductors.
Although weak-coupling theories explain the experimental phase diagram qualitatively, a quantitative discrepancy between them has discouraged conclusive interpretation.
In this Letter, we thoroughly study the effect of Coulomb interaction and the phase diagrams of locally noncentrosymmetric superconductors.
We reveal even-odd parity transition and the enhancement of the parity transition field in strongly correlated superconductors, and an issue of CeRh$_2$As$_2$ is resolved.
\end{abstract}

\maketitle

\textit{Introduction.} --- 
Searching for odd-parity superconductors has been a central issue in designing topological materials~\cite{Qi2011,Tanaka2012,Sato2016,Sato2017}.
Due to Fermi statistics, odd-parity superconductors are classified into spin-triplet superconductors within the theory by Bardeen, Cooper, and Schrieffer (BCS)~\cite{Bardeen1957}, which brilliantly explains various phenomena of superconductivity.
In the topological science, this constraint on Cooper pairs has led to the renewed interest in uncommon spin-triplet superconductors such as UPt$_3$~\cite{Joynt2002,Tsutsumi2013,Yanase2016,Yanase2017}, UCoGe~\cite{Aoki2014,Aoki2019,Daido2019}, and UTe$_2$~\cite{Ishizuka2019,Ishizuka2020,Aoki2022,Fujibayashi2022}.
However, some internal degrees of freedom of Cooper pairs are overlooked in the canonical BCS theory.

Recently, the local noncentrosymmetricity in superconductors has attracted much attention and shed light on the sublattice degree of freedom in Cooper pairs~\cite{Fischer2011,Maruyama2012,Yoshida2012,Yoshida2013,Yoshida2014,Yoshida2015,Higashi2016,Shimozawa2016,Wu2017,Yoshida2017,Nakamura2017,Gotlieb2018,Mockli2018,Skurativska2021,Fischer2022}. 
Interestingly, the sublattice antisymmetric Cooper pair is allowed, leading to an odd-parity superconducting state without spin-triplet pairs.
Therefore, the sublattice degree of freedom paves a new way to design topological odd-parity superconductors based on pairing in the ordinary spin-singlet channel~\cite{Yoshida2015,Nogaki2021}.
In the high-magnetic field phase of locally noncentrosymmetric superconductors, the sublattice antisymmetric superconducting state is theoretically predicted to be thermodynamically stable and named the pair-density-wave (PDW) state~\cite{Yoshida2012}.

The discovery of the two superconducting phases in the $H$-$T$ phase diagram of CeRh$_2$As$_2$~\cite{Khim2021} has led attention to the sublattice degree of freedom in superconductors~\cite{Khim2021,Pourret2021,Kimura2021,Onishi2022,Hafner2022,Kibune2022,Kitagawa2022,Landaeta2022,Schertenleib2021,Mockli2021,Ptok2021,Nogaki2021,Mockli2021_2,Cavanagh2022,Hazra2022}.
In fact, due to the fascinating crystalline structure of CeRh$_2$As$_2$, the inversion symmetry is locally broken at the Ce site but globally preserved~\cite{Khim2021}.
The qualitative similarity of the phase diagrams between the weak-coupling theory~\cite{Yoshida2012,Khim2021} and experiment~\cite{Khim2021,Landaeta2022} strongly suggests an essential role of local inversion symmetry breaking in CeRh$_2$As$_2$, and the two superconducting phases were interpreted 
based on the even-odd parity transition within the superconducting state~\cite{Khim2021,Landaeta2022}.

In contrast to the preceding argument, there are two issues regarding the nature of CeRh$_2$As$_2$. 
First, the microscopic mechanism of superconductivity in CeRh$_2$As$_2$ has been unsolved.
The unconventional superconductivity mediated by quantum critical fluctuations is studied in this Letter. 
Second, the parity transition field of CeRh$_2$As$_2$ significantly exceeds the Pauli-Clogston-Chandrasekhar limit and is larger than the prediction of weak-coupling theory by a factor of five~\cite{Yoshida2012,Khim2021}.
Although two scenarios have been proposed within the weak-coupling theory~\cite{Cavanagh2022,Mockli2021_2}, 
we try to resolve the issue by verifying an intrinsic phase diagram of strongly correlated locally noncentrosymmetric superconductors.

To tackle above mentioned problems, we focus on the electronic correlation effect of Ce $f$-electrons, which have localized character.
Indeed, the large electronic specific heat coefficient $\gamma \sim 1000\,\mathrm{mJ/mol\,K^2}$ supports the presence of heavy-fermion bands near the Fermi-level, and non-Fermi liquid behaviors suggest quantum criticality in CeRh$_2$As$_2$~\cite{Khim2021,Hafner2022,Kitagawa2022}. 
These experimental observations indicate that Coulomb interaction crucially impacts the electronic state of CeRh$_2$As$_2$.
Hence, theoretical studies of strong correlation effects in locally noncentrosymmetric superconductors have been desired.

In this Letter, we conduct a thorough study on quantum critical multipole fluctuations, resulting superconductivity, and superconducting phase diagrams of locally noncentrosymmetric strongly correlated electron systems.
To clarify these properties, fluctuation exchange (FLEX) approximation which appropriately reproduces critical behaviors of self-consistent renormalization theory~\cite{Moriya2000} is adopted.
In the FLEX scheme, the Green function and the self-energy depend on each other, and therefore, the retardation effect, quasi-particle scattering, and internal field are taken into account~\cite{1supplement}.
Theoretical results are compared with the superconducting phase diagrams of CeRh$_2$As$_2$, and the origin of superconductivity is discussed.
As a result, the $XY$-type antiferromagnetic fluctuation consistent with the nuclear magnetic resonance (NMR)~\cite{Kitagawa2022} and superconductivity with dominant $d_{x^2-y^2}$-wave pairing are revealed, and the obtained enhanced parity transition field resolves the issue of phase diagram in CeRh$_2$As$_2$.

\textit{Bilayer Rashba-Hubbard model.} --- 
\begin{figure}[tbp]
 \begin{center}
    \includegraphics[keepaspectratio, scale=0.25]{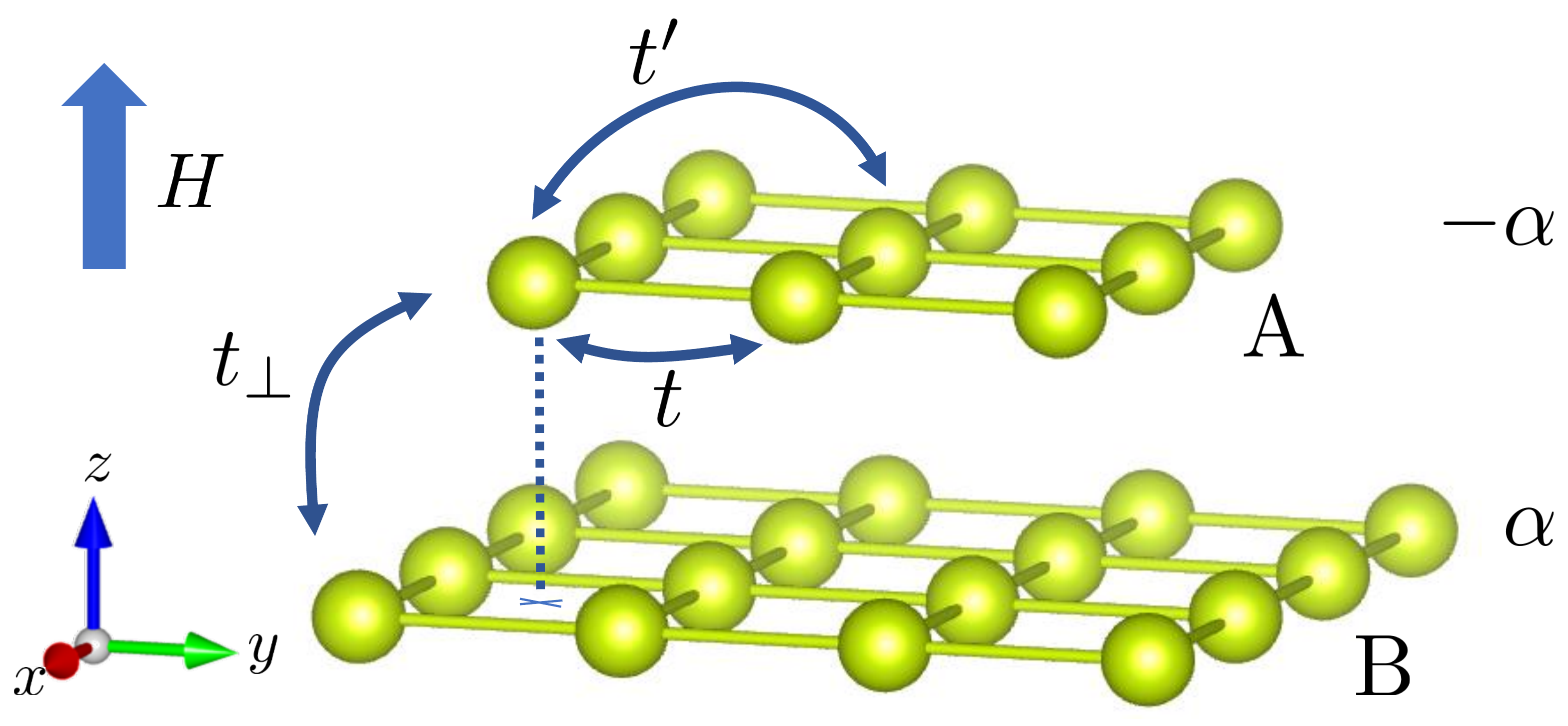}
  \end{center}
  \caption{Schematic figure of the bilayer Rashba-Hubbard model.
  Yellow circles represent the Ce atoms of CeRh$_2$As$_2$.
  We introduce first- and second-neighbor intra-layer hopping integrals.
  We also introduce an inter-layer hopping integral as $t_{\perp}$.
  The staggered Rashba-type antisymmetric spin-orbit coupling is included in the model as it arises from the asymmetric potential by Rh$_2$As$_2$ layers.
  The magnetic field is applied parallel to the $z$-axis.}
  \label{fig:model}
\end{figure}
We construct the bilayer Rashba-Hubbard model, in which Ce $f$-orbital, Coulomb correlation, and spin-orbit coupling are taken into account~\cite{Craco2021,Lu2021}.
The model is given by
\begin{align}
  \hat{H} = \sum_{\bk} \varphi^\dagger(\bk) \mathcal{H}_0(\bk) \varphi(\bk) + U \sum_{i,\sigma} n_{i\uparrow \sigma}n_{i\downarrow \sigma},
\end{align}
where $U$ is the on-site Coulomb repulsion, $\mathcal{H}_0(\bk) = \varepsilon(\bm{k}) s_0\otimes\sigma_0 + \alpha \bm{g}(\bm{k}) \cdot \bm{s} \otimes \sigma_z -\mu_B H s_z \otimes \sigma_0 + \tilde{t}_{\perp}(\bk) s_0\otimes\sigma_+ + \tilde{t}_{\perp}(-\bk) s_0\otimes\sigma_-$, $\varphi(\bk) = (c_{\bk \uparrow \mathrm{A}}, c_{\bk \downarrow \mathrm{A}}, c_{\bk \uparrow \mathrm{B}},c_{\bk \downarrow \mathrm{B}})^{\top}$, and $c_{\bm{k} s \sigma}$ ($c^{\dagger}_{\bm{k} s \sigma}$) is an annihilation (creation) operator for an electron with momentum $\bm{k}$, spin $s$, and sublattice $\sigma \in \{\mathrm{A},\mathrm{B}\}$ [Fig.~\ref{fig:model}]. 
Here, $s_\mu$ and $\sigma_\mu$ consisting of the $2\times2$ unit matrix and three Pauli matrices represent spin and sublattice degrees of freedom, respectively.\footnote{$\mu$ runs over $\{0,x,y,z\}$, and $\sigma_{\pm}$ are given by $(\sigma_x\pm i \sigma_y)/2$.}
The first term of $\mathcal{H}_0$ represents intra-layer hopping including the chemical potential and is given by $\varepsilon(\bm{k}) = -2t(\cos k_x+\cos k_y)+4t'\cos k_x \cos k_y -\mu$. 
The vector $\bm{g}(\bm{k})$ describes the Rashba-type antisymmetric spin-orbit coupling given by $\bm{g}(\bm{k}) = (-\partial \varepsilon(\bm{k})/\partial k_y ,\partial \varepsilon(\bm{k})/\partial k_x,0)$, and $H$ represents the Zeeman magnetic field parallel to the $z$-axis. 
The last two terms of $\mathcal{H}_0$ describe inter-layer hopping given as $\tilde{t}_{\perp}(\bk)=t_{\perp}(1+e^{-ik_x})(1+e^{-ik_y})$.
This model is a straightforward extension of the Rashba-Hubbard model for globally inversion-asymmetric strongly correlated electron systems~\cite{Bauer2012,Shigeta2013,Fujimoto2015,Maruyama2015,Greco2018,Lu2018,Ghadimi2019,Greco2020,Nogaki2020,Wolf2020,Trott2020,Soni2021,Biderang2022}. 
Hereafter, we set $t'=0.3$, $t_{\perp}=0.1$, $\mu_{\mathrm{B}}=1$, and $U=3.9$ with a unit of energy $t=1$ and determine the chemical potential so that the electron density per site $n$ is $0.85$.

\textit{Multipole susceptibility.} --- 
First, we discuss quantum critical multipole fluctuations. 
Dynamical susceptibility tensor is given by the generalized susceptibility as
\begin{align}
  \chi_{\hat{\mathcal{O}}}(\bm{q},i\nu_n)=\sum_{\xi_1\xi_2\xi_3\xi_4}\hat{\mathcal{O}}_{\xi_1\xi_2}\chi_{\xi_2\xi_1\xi_3\xi_4}(\bm{q},i\nu_n)\hat{\mathcal{O}}_{\xi_3\xi_4},
\end{align}
where $i\nu_n$ are bosonic Matsubara frequencies and $\xi$ is abbreviated notation $\xi=(s,\sigma)$.
The operators $\hat{\mathcal{O}}$ are (extended) multipole operators as $\hat{\mathcal{O}} = \hat{s}\otimes \hat{\sigma}$~\cite{Watanabe2018,Hayami2018,Yatsushiro2021}, 
and classified into even-parity (odd-parity) multipole for $\hat{\sigma}_0, \hat{\sigma}_x$ ($\hat{\sigma}_y, \hat{\sigma}_z$).
In Table~\ref{tab:multi}, we summarize the classification of multipole operators in our system.
We adopt the normalization convention  $\mathrm{tr}\left[\hat{\mathcal{O}}^\dagger\hat{\mathcal{O}}\right]=1$, and therefore, operators are represented by the Pauli matrices as $\hat{s}_\mu=s_\mu/\sqrt{2}$ and $\hat{\sigma}_\mu=\sigma_\mu/\sqrt{2}$.

\begin{figure}[tbp]
 \begin{center}
    \includegraphics[keepaspectratio, scale=0.45]{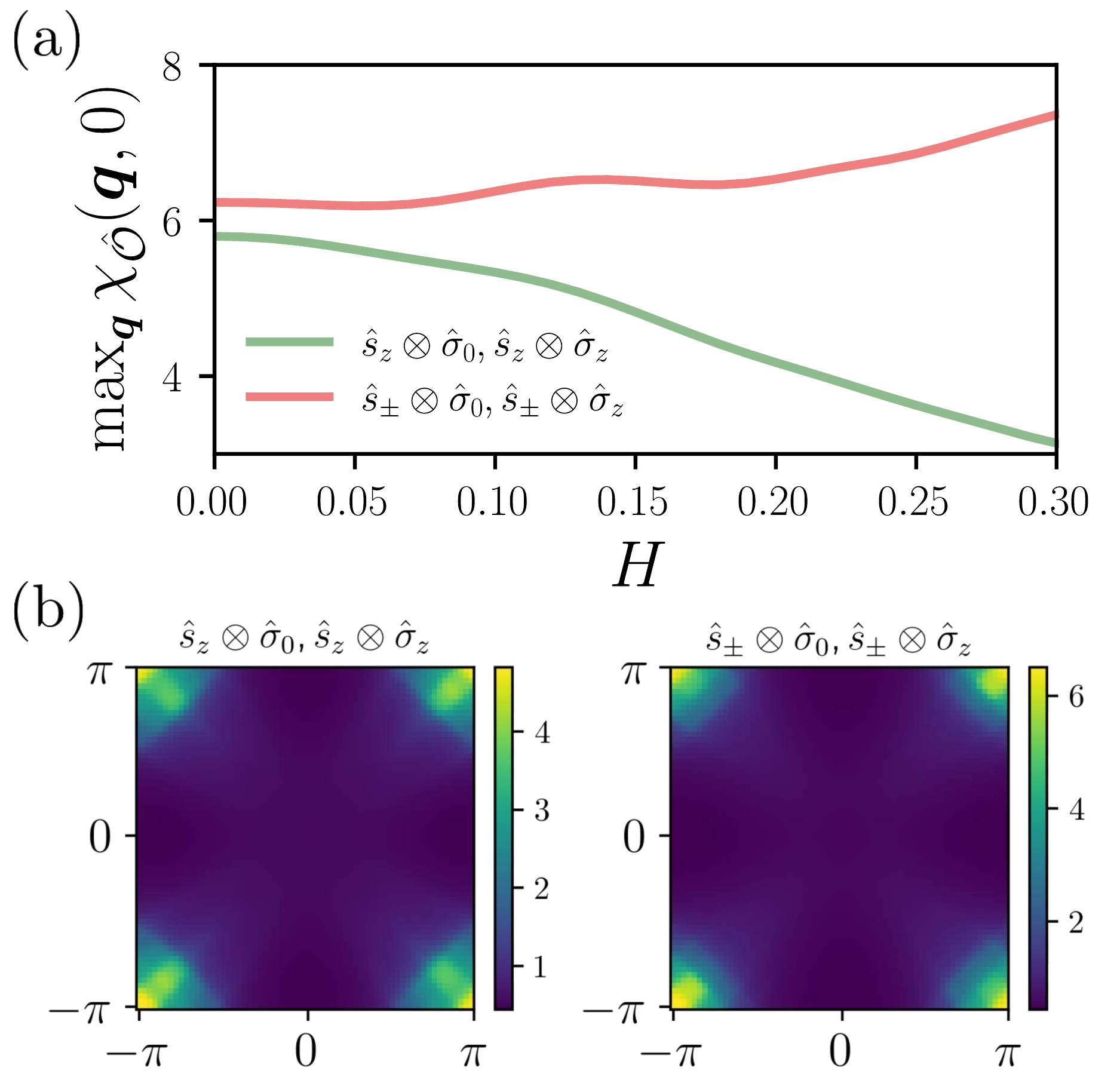}
  \end{center}
  \caption{(a) The magnetic field dependence of static multipole fluctuations. 
  The maxima of the longitudinal and transverse magnetic susceptibilities are shown. 
  Note that the even-parity and odd-parity multipole fluctuations are nearly degenerate. 
  We assume $\alpha/t_{\perp}=2$ and $T=0.01$.
  (b) The momentum dependence of longitudinal magnetic susceptibility (left) and transverse magnetic susceptibility (right) at $H=0.15$.}
  \label{fig:suscep}
\end{figure}

In Fig.~\ref{fig:suscep}(a), the magnetic field dependence of the maximum of transverse magnetic and longitudinal magnetic susceptibilities is shown.
We see that the transverse (longitudinal) susceptibility is enhanced (reduced) by the magnetic field. 
Note that the even-parity magnetic multipole fluctuation for $\bm{s} \otimes \sigma_{0}$ and the odd-parity magnetic multipole fluctuation for $\bm{s} \otimes \sigma_{z}$ are nearly degenerate, while the odd-parity multipole is slightly favored.
Other multipole susceptibilities are negligibly small.
From Fig.~\ref{fig:suscep}(b), both transverse and longitudinal magnetic susceptibilities show the peak structure around $\bm{Q}=(\pi,\pi)$, and thus, the antiferromagnetic spin fluctuation develops in the system.
The momentum dependence is not qualitatively affected by the magnetic field.
Therefore, we conclude dominant $XY$-type antiferromagnetic fluctuation, which is consistent with the observation in CeRh$_2$As$_2$ by the NMR measurement~\cite{Kitagawa2022}.
We confirmed that the behaviors of the multipole susceptibility do not qualitatively depend on the strength of spin-orbit coupling $\alpha/t_{\perp}$~\cite{2supplement}.

\begin{table}[htb]
  \begin{center}
  \caption{The classification of the multipole operators in the bilayer Rashba-Hubbard model.
  Here, $\hat{s}_{\pm}=(\hat{s}_x \pm i\hat{s}_y)/\sqrt{2}$ are ladder operators for spin.
  E (O) represents the even-parity (odd-parity) multipole operators. 
  C, L, and T represent charge, longitudinal spin, and transverse spin operators respectively. 
  $\hat{\sigma}_0$ and $\hat{\sigma}_z$ ($\hat{\sigma}_x$ and $\hat{\sigma}_y$) are intra-sublattice (inter-sublattice) operators.}
\begin{tabular}{|c|c|c|c|c|}
\hline 
$\hat{\mathcal{O}}$ & $\hat{\sigma}_0$ & $\hat{\sigma}_x$ & $\hat{\sigma}_y$ & $\hat{\sigma}_z$ \\
\hline 
$\hat{s}_0$ & E C intra & E C inter & O C intra & O C inter \\
$\hat{s}_z$ & E L intra & E L inter & O L intra & O L inter \\
$\hat{s}_{\pm}$ & E T intra & E T inter & O T intra & O T inter \\
\hline  
  \end{tabular}
  \label{tab:multi}
  \end{center}
\end{table}

\textit{Superconductivity.} --- 
The crystalline space group of CeRh$_2$As$_2$ is $P4/nmm$ (No.129), and therefore, the point group is $D_{4h}$.
Since we introduce the magnetic field parallel to the $z$-axis, some symmetry operations are prohibited, and the point group reduces to $C_{4h}$.
Hence, the superconducting gap functions are classified based on irreducible representations of $C_{4h}$.
Using the conventional notation, we decompose the superconducting gap functions into the spin-singlet component and spin-triplet component, 
\begin{align}
    \Delta^{\sigma\sigma'}(\bk) = \{\psi^{\sigma\sigma'}(\bk)+\bm{d}^{\sigma\sigma'}(\bk)\cdot\bm{s}\} is_y,
\end{align}
for the intra-sublattice and inter-sublattice pairing channels. Here, $\psi^{\mathrm{AA}}(\bk)$ ($\psi^{\mathrm{AB}}(\bk)$) represents an intra-sublattice (inter-sublattice) spin-singlet order parameter, while $\bm{d}^{\mathrm{AA}}(\bk)$ ($\bm{d}^{\mathrm{AB}}(\bk)$) is an intra-sublattice (inter-sublattice) spin-triplet order parameter.
In the following calculations, 
order parameters of inter-sublattice pairing are negligibly small. 
The basis functions of intra-sublattice order parameter for each irreducible representations of $C_{4h}$ are summarized in Table~\ref{tab:basis}.

\begin{figure}[tbp]
 \begin{center}
    \includegraphics[keepaspectratio, scale=0.48]{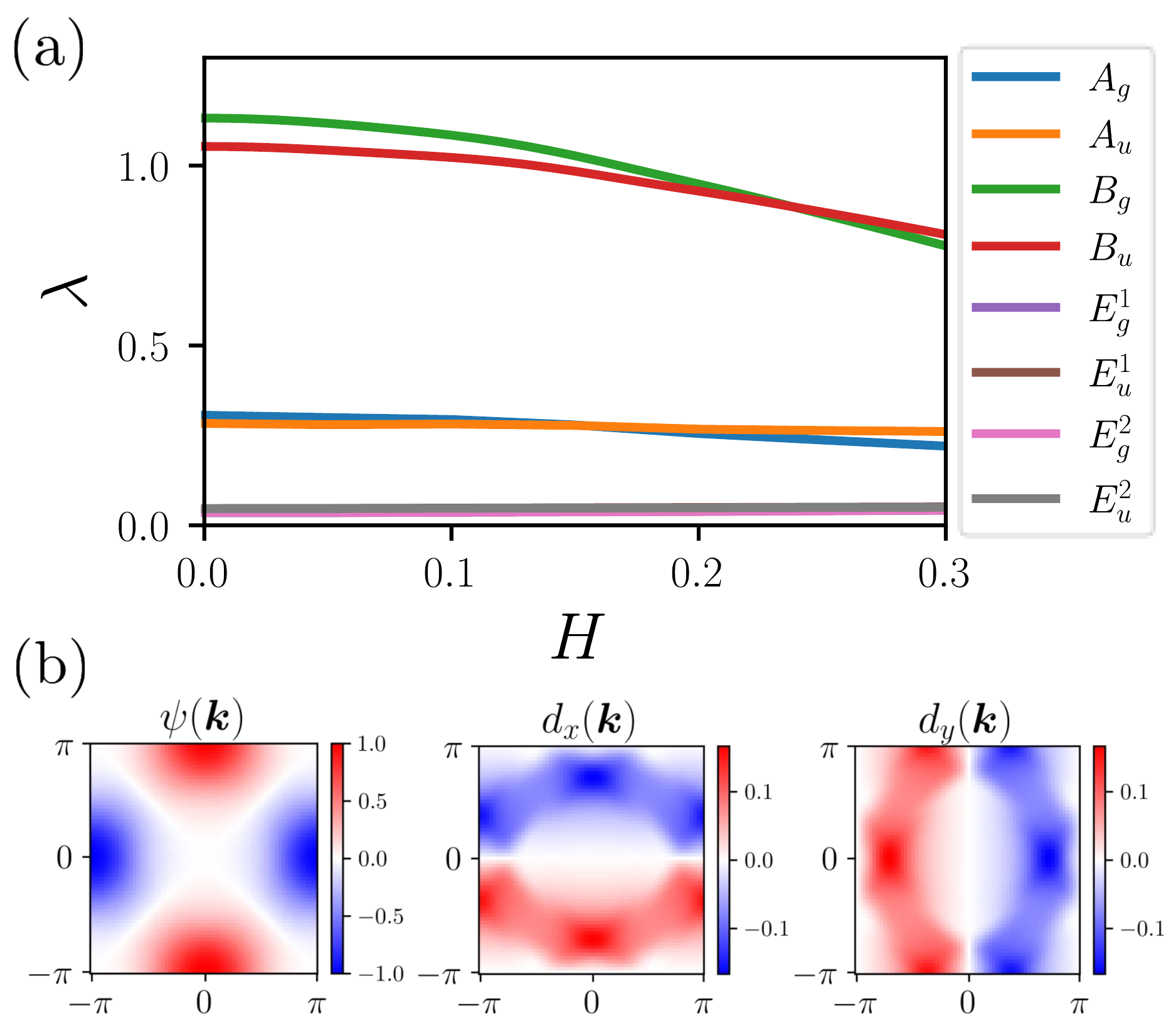}
  \end{center}
  \caption{ 
  (a) The magnetic field dependence of eigenvalues of the \'{E}liashberg equation for each irreducible representation. 
  We assume $\alpha/t_{\perp}=2$ and $T=0.01$.
  (b) The momentum dependence of intra-sublattice spin-singlet and spin-triplet gap functions, $\psi^{\mathrm{AA}}(\bk)$ and $\bm{d}^{\mathrm{AA}}(\bk)$, of the $B_g$ representation for $H=0.15$. 
  Note that $d_z^{\mathrm{\sigma\sigma}}(\bk)=0$. Results for the $B_u$ representation are almost the same as the figures.}
  \label{fig:delta}
\end{figure}

In Fig.~\ref{fig:delta}(a), the eigenvalues of \'{E}liashberg equation for each irreducible representation are shown~\cite{3supplement}.
The $B_g$ and $B_u$ representations are dominant.
From the momentum dependence displayed in Fig.~\ref{fig:delta}(b), both of these states contain spin-singlet $d_{x^2-y^2}$-wave dominant pairing as well as spin-triplet subdominant pairing with $p$-wave symmetry.
These unconventional Cooper pairs are stabilized by the antiferromagnetic fluctuations and spin-orbit coupling and not significantly changed against the magnetic field.

While almost the same momentum dependence, the different sublattice structures of gap functions distinguish the $B_g$ and $B_u$ representations.
In the even-parity $B_g$ representation, the spin-singlet (spin-triplet) gap function has the same (opposite) sign between the sublattices A and B, as $\psi^{\mathrm{AA}}(\bk)=\psi^{\mathrm{BB}}(\bk)$ and $\bm{d}^{\mathrm{AA}}(\bk)=-\bm{d}^{\mathrm{BB}}(\bk)$.
On the other hand, $\psi^{\mathrm{AA}}(\bk)=-\psi^{\mathrm{BB}}(\bk)$ and $\bm{d}^{\mathrm{AA}}(\bk)=\bm{d}^{\mathrm{BB}}(\bk)$ 
in the odd-parity $B_u$ representation.
Therefore, the $B_g$ and $B_u$ representations correspond to the BCS and PDW states predicted in the weak-coupling theory~\cite{Yoshida2012}.
As shown in Fig~\ref{fig:delta}(a), eigenvalues of the \'{E}liashberg equation for both $B_g$ and $B_u$ representations are weakened by the magnetic field due to the Pauli depairing effect.
However, the $B_u$ state is more robust compared with the $B_g$ state, and therefore, at $H=0.24$ the parity transition from the even-parity $B_g$ state to the odd-parity $B_u$ state occurs.
The transition can be understood from the viewpoint of the intrinsic magnetic response of these states. 
Indeed, the $B_g$ state is Pauli-limited, but the $B_u$ state mostly avoids the Pauli limiting because the magnetic susceptibility does not decrease through the superconducting transition~\cite{Maruyama2012,Skurativska2021}.

\begin{table}[htb]
  \begin{center}
  \caption{The basis functions for intra-sublattice superconducting order parameter. IR represents the irreducible representations of the point group $C_{4h}$. We take into account the time-reversal symmetry breaking under the magnetic field, and the degeneracy of the $E_{g/u}$ states is lifted. Thus, we distinguish them as $E^1_{g/u}$ and $E^2_{g/u}$. Note that the spin-singlet component $\psi(\bk)$ and spin-triplet in-plane component $d_{x,y}(\bk)$ for the $E$ representations are prohibited due to the $C^z_2$ rotation symmetry.
  }
\begin{tabular}{|c|c|c|}
\hline 
IR & $\psi(\boldsymbol{k})$ & $\bm{d}(\bm{k})$ \\
\hline \hline 
$A_{g}$, $A_{u}$ & $1$, $k_{x} k_{y}\left(k^2_{x}-k^2_{y}\right)$ & $k_x\hat{\boldsymbol{x}}+k_{y} \hat{\boldsymbol{y}}$, $k_{y} \hat{\boldsymbol{x}}-k_{x} \hat{\boldsymbol{y}}$ \\
$B_{g}$, $B_{u}$ & $k_{x}k_{y}$, $k^2_{x}-k^2_{y}$& $k_{x} \hat{\boldsymbol{x}}-k_{y} \hat{\boldsymbol{y}}$, $k_{y} \hat{\boldsymbol{x}}+k_{x} \hat{\boldsymbol{y}}$ \\
$E^1_{g}$, $E^1_{u}$ &  0 & $\left(k_{x}+ik_{y}\right) \hat{\boldsymbol{z}}$ \\
$E^2_{g}$, $E^2_{u}$ & 0 & $\left(k_{x}-ik_{y}\right) \hat{\boldsymbol{z}}$ \\
\hline  
  \end{tabular}
  \label{tab:basis}
  \end{center}
\end{table}

\textit{Phase diagram.} --- 
\begin{figure*}[t]
 \begin{center}
    \includegraphics[keepaspectratio, scale=0.5]{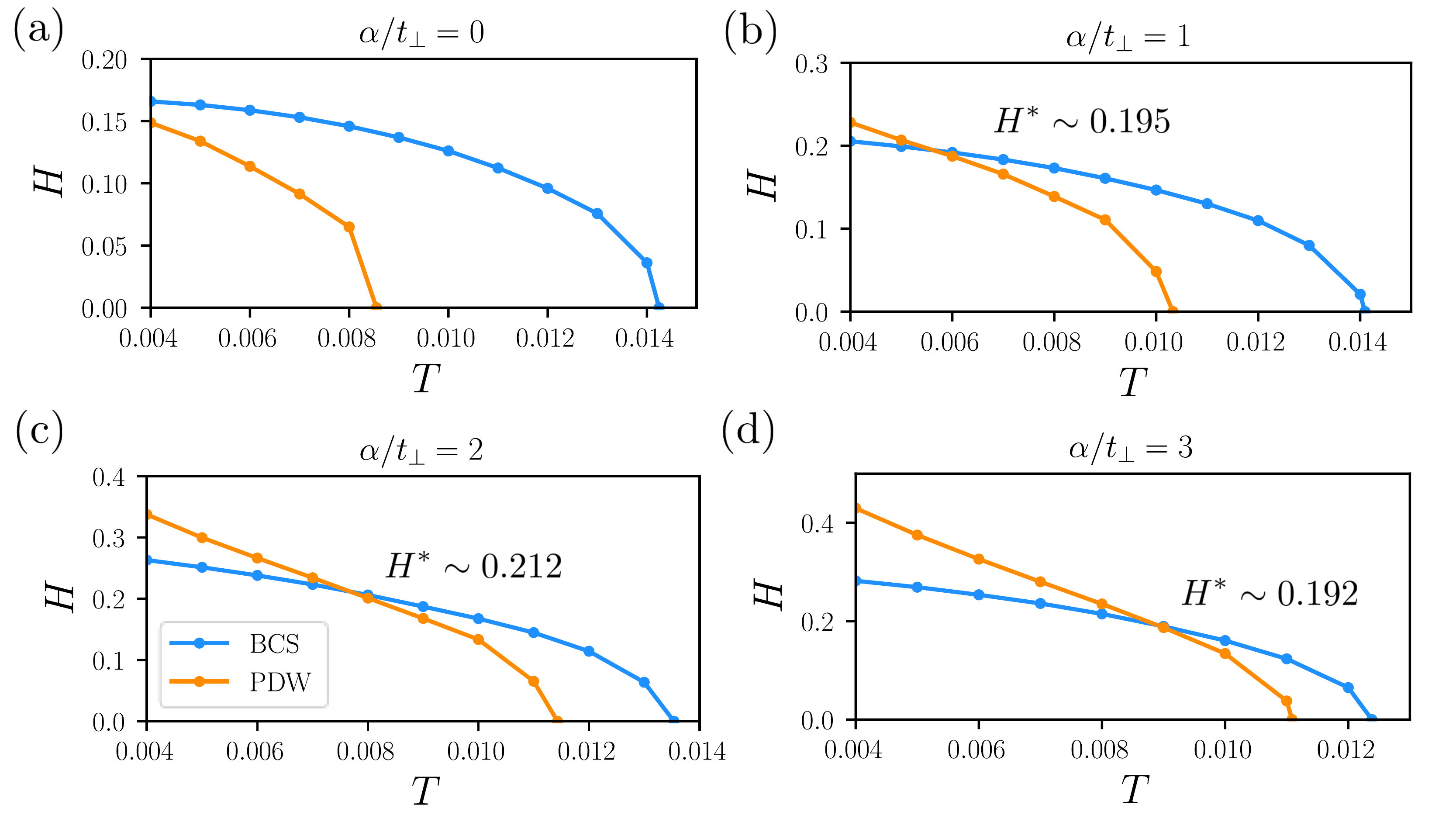}
  \end{center}
  \caption{(a)-(d) $H$-$T$ phase diagrams of the bilayer Rashba-Hubbard model for $\alpha/t_{\perp}=0,1,2,3$.
  We show the superconducting transition lines of the even-parity $B_{g}$ and odd-parity $B_{u}$ states, on which eigenvalues of the \'{E}liashberg equation become unity.
  The BCS and PDW states correspond to the $B_{g}$ and $B_{u}$ superconducting phases, respectively.
  $H^*$ denotes the magnetic field at the parity transition point.
  }
  \label{fig:phases}
\end{figure*}
Let us discuss the $H$-$T$ phase diagram of the bilayer Rashba-Hubbard model.
In Figs.~\ref{fig:phases}(a)-(d), we show the phase diagrams for $\alpha/t_{\perp}=0$, $1$, $2$, and $3$, which is known as a control parameter of local noncentrosymmetricity~\cite{Maruyama2012}.\footnote{While the case $\alpha/t_{\perp}=0$ corresponds to the bilayer system without spin-orbit coupling, for $\alpha/t_{\perp}=\infty$ the system is equivalent to a set of monolayer systems with Rashba-type spin-orbit coupling~\cite{Maruyama2012}.}
We show the superconducting transition lines~\cite{4supplement} 
of the $B_{g}$ and $B_{u}$ states, which correspond to the BCS and PDW states as mentioned before.

From the obtained phase diagrams, the zero-field superconducting transition temperature $T_{\rm c}$ and the magnetic field at the parity transition point $H^*$ are estimated as 
$(T_{\rm c},H^*)=(0.0141,0.195)$,  $(0.0135,0.212)$, and $(0.0124,0.192)$ for 
$\alpha/t_{\perp}=1$, $2$, and $3$, respectively.
From these estimations, we evaluate $H^*$ with a unit of $T_{\rm c}$,
\begin{align}
\frac{H^*}{T_{\rm c}} =
\begin{cases}
13.8 & (\alpha/t_{\perp}=1) \\
15.7 & (\alpha/t_{\perp}=2) \\
15.5 & (\alpha/t_{\perp}=3).
\end{cases}
\label{eq:ratio}
\end{align}
Thus, we conclude that the parity transition fields scaled by transition temperature are approximately $H^*/T_{\rm c} \simeq 15$ and universal against variation of $\alpha/t_{\perp}$.
Although a much smaller value $H^*/T_{\rm c} \simeq 2$ was predicted by the mean-field theory~\cite{Yoshida2012}, 
in CeRh$_2$As$_2$ 
experimental values $H^* \simeq 3.9$\,T and $T_{\rm c} \simeq 0.26$\,K 
lead $(H^*/T_{\rm c})_{\mathrm{exp}} \simeq 10$~\cite{Khim2021}.
Hence, our theoretical result is quantitatively consistent with the phase diagram of CeRh$_2$As$_2$ 
and resolves the discrepancy between the weak-coupling theory and experiment.

\begin{figure}[tbp]
 \begin{center}
    \includegraphics[keepaspectratio, scale=0.25]{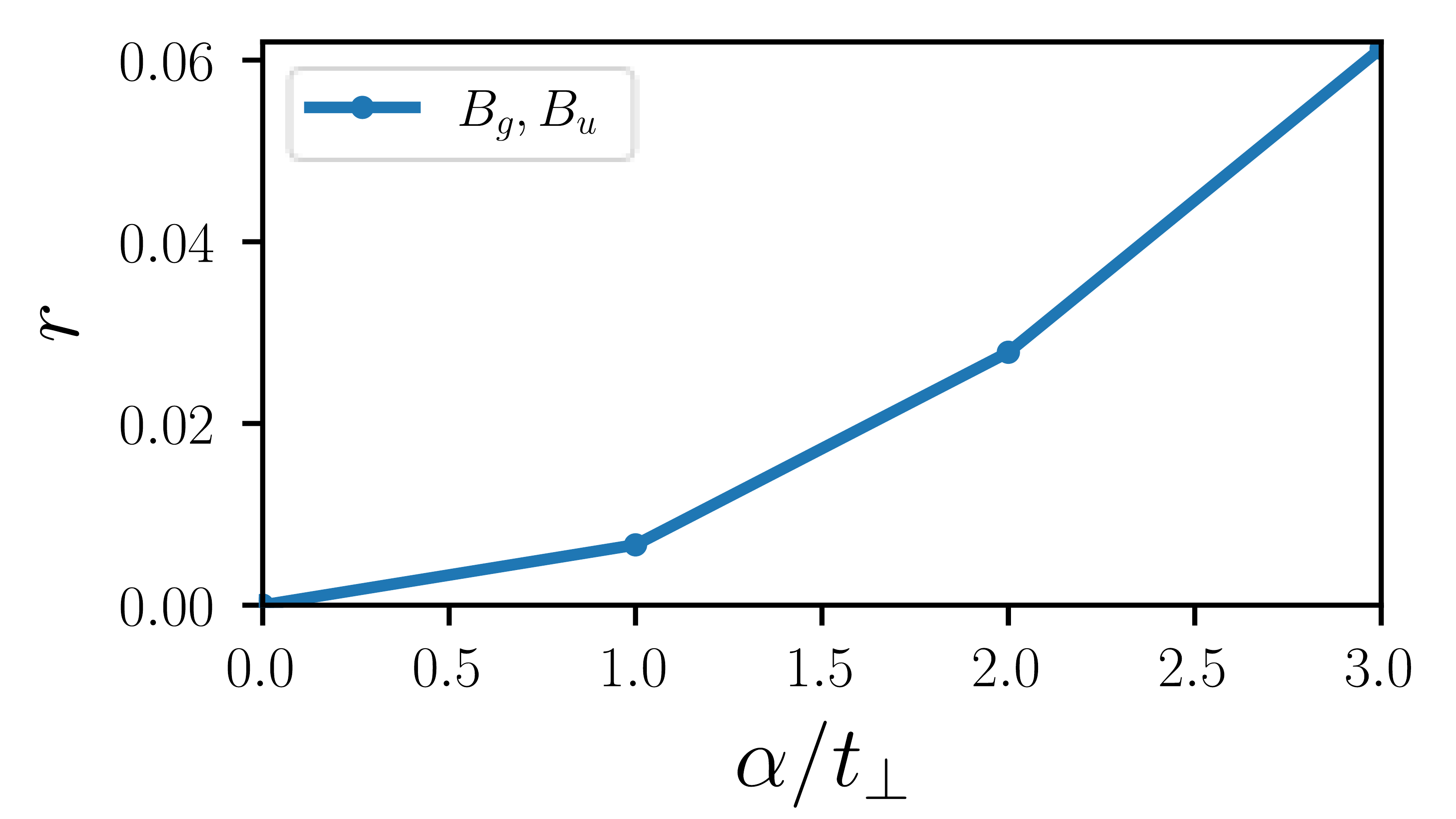}
  \end{center}
  \caption{ 
  The parity mixing parameter $r$ as a function of $\alpha/t_{\perp}$.
  The values for the $B_g$ and $B_u$ representations are nearly the same.
  We assume $T=0.01$ and $H=0.15$.}
  \label{fig:parity_ratio}
\end{figure}

\begin{figure}[tbp]
 \begin{center}
    \includegraphics[keepaspectratio, scale=0.25]{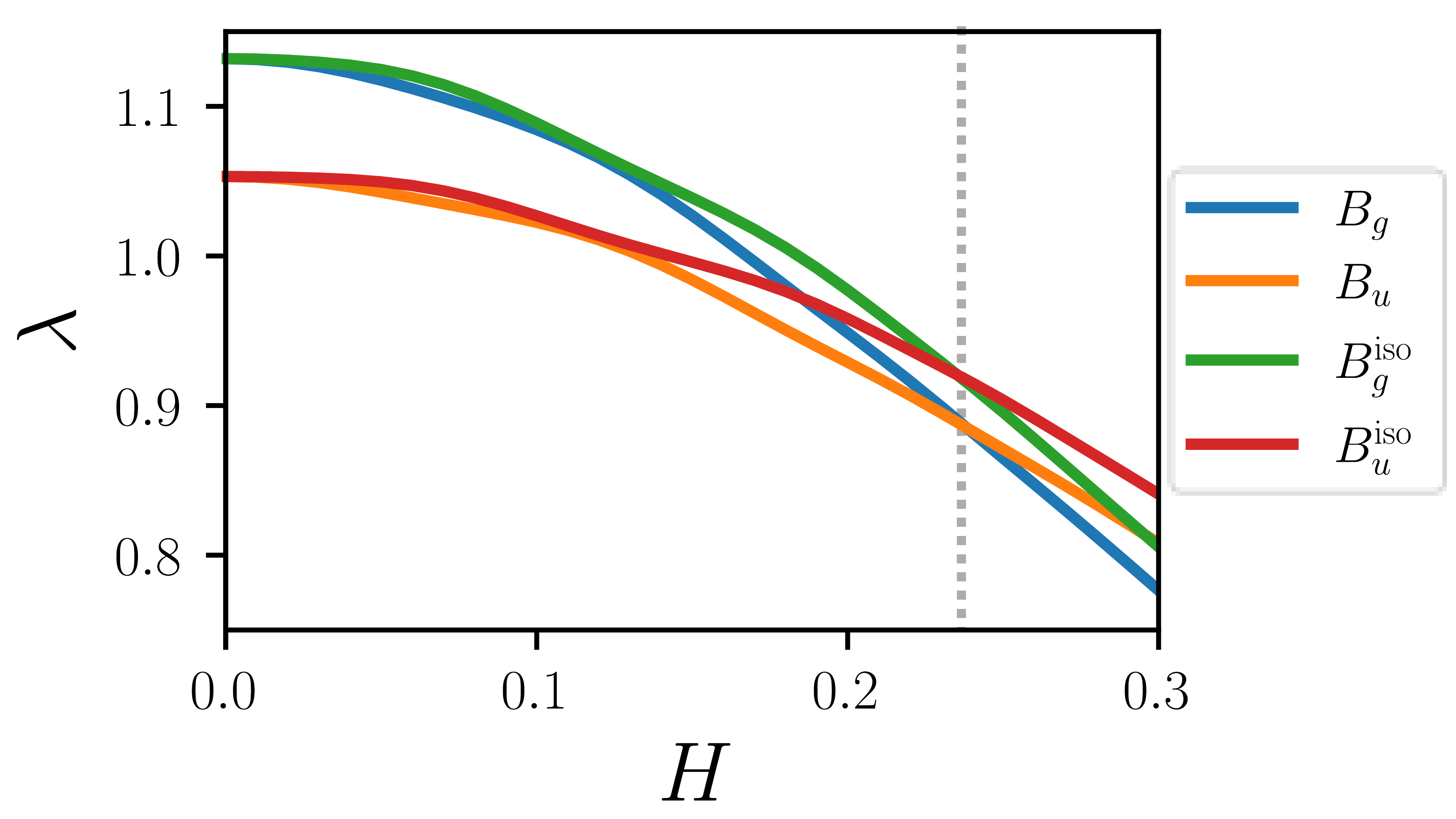}
  \end{center}
  \caption{ 
  The magnetic field dependence of eigenvalues of the \'{E}liashberg equation for the $B_g$ and $B_u$ representations. 
  $B^{\mathrm{iso}}_g$ and $B^{\mathrm{iso}}_u$ represent the eigenvalues without the field-enhanced magnetic anisotropy.
  The parity transition points are indicated by the gray dashed line.
  We assume $\alpha/t_{\perp}=2$ and $T=0.01$.}
  \label{fig:iso_delta}
\end{figure}

Now we discuss the origin of the enhancement in the parity transition field $H^*$. First, a possible origin is the nonsymmorphic crystalline symmetry as proposed in Ref.~\onlinecite{Cavanagh2022}. Actually, an indicator of local noncentrosymmetricity $\alpha/\tilde{t}_{\perp}(\bk)$ diverges at the Brillouin Zone faces in nonsymmorphic crystals because of $\tilde{t}_{\perp}(\bk_{\rm face})=0$, and therefore, 
the effect of the spin-orbit coupling is more essential than the symmorphic case~\cite{Sumita2016,Cavanagh2022}. 
However, our analysis of the bilayer Rashba-Hubbard model does not support this possibility, because the ratio $H^*/T_{\rm c}$ is universal against variation of $\alpha/t_{\perp}$. 
Second, we may expect that the spin-triplet component in the gap function changes the spin state of Cooper pairs and increases $H^*$. 
However, this is also unlikely because 
we observe strong dependence on $\alpha/t_{\perp}$ of the parity mixing parameter $r$ (Fig.~\ref{fig:parity_ratio}), which is defined as
\begin{align}
    r = \frac{\max_{\bk}|\bm{d}^{\mathrm{AA}}(\bk)|^2}{\max_{\bk}|\psi^{\mathrm{AA}}(\bk)|^2}.
\end{align}
From Eq.~\eqref{eq:ratio} and Fig.~\ref{fig:parity_ratio}, we find that $H^*/T_{\rm c}$ and $r$ are almost uncorrelated.
Third, we rule out the field dependence of effective interaction as the main origin. 
In the bilayer Rashba-Hubbard model, the effective pairing interaction is field-dependent because of the field-enhanced magnetic anisotropy (see Fig.~\ref{fig:suscep}).
Figure~\ref{fig:iso_delta} shows eigenvalues of the \'{E}liashberg equation with and without the field dependence of magnetic anisotropy,\footnote{To investigate the effect of the field-induced change of effective interaction, we solve the \'{E}liashberg equation with the zero-field vertex functions while the Green functions at $H\ne 0$ are adopted.}
and we find that the field-enhanced magnetic anisotropy weakens the superconducting instabilities. 
However, the parity transition field $H^*$ is hardly affected. 
Thus, the field-enhanced magnetic anisotropy is also irrelevant for the enhancement of $H^*$.
We conclude from these considerations that 
a large parity transition field consistent with CeRh$_2$As$_2$ is 
due to the internal field arising from the quantum critical antiferromagnetic fluctuation. Near the antiferromagnetic critical point, the spin correlation significantly screens the Zeeman field and increases the scale of magnetic fields~\cite{Yanase2008}. This is consistent with the large upper critical fields of CeRh$_2$As$_2$ which significantly exceeds the Pauli-Clogston-Chandrasekhar limit even for the in-plane direction~\cite{Khim2021}.

\textit{Conclusion.} --- We investigated the nature of quantum critical multipole fluctuation and superconductivity in the bilayer Rashba-Hubbard model, a minimal model of locally noncentrosymmetric strongly correlated electron systems. Experimental observations in a recently discovered superconductor CeRh$_2$As$_2$~\cite{Khim2021} were discussed. 
The $XY$-type antiferromagnetic fluctuations are shown consistent with the NMR study~\cite{Kitagawa2022}, and the applied magnetic field along the $z$-axis enhances the fluctuations.
Due to the critical antiferromagnetic fluctuation, superconductivity with dominant $d_{x^2-y^2}$-wave pairing and subdominant $p$-wave pairing is stabilized, and the two superconducting phases with different space inversion parity appear in the $H$-$T$ phase diagram, consistent with CeRh$_2$As$_2$. 
The parity transition field is enhanced by the quantum critical fluctuation, and the obtained value is in quantitative agreement with the experimental value.
These results support the parity transition in the superconducting state of CeRh$_2$As$_2$~\cite{Khim2021} and indicate the topological superconductivity~\cite{Nogaki2021}.
Our theory not only solves the issues of CeRh$_2$As$_2$ but also elucidates general behaviors of the family of locally noncentrosymmetric strongly correlated superconductors.

\begin{acknowledgments}
The authors are grateful to A.~Daido and S.~Sumita for fruitful discussions.
Some figures in this work were created by using {\sc Vesta}~\cite{Momma2011}.
This work was supported by JSPS KAKENHI (Grants Nos. JP18H01178, JP18H05227, JP20H05159, JP21J23007, JP21K18145, JP22H01181) and SPIRITS 2020 of Kyoto University.
\end{acknowledgments}

\bibliography{paper}

\clearpage

\renewcommand{\bibnumfmt}[1]{[S#1]}
\renewcommand{\citenumfont}[1]{S#1}
\renewcommand{\thesection}{S\arabic{section}}
\renewcommand{\theequation}{S\arabic{equation}}
\setcounter{equation}{0}
\renewcommand{\thefigure}{S\arabic{figure}}
\setcounter{figure}{0}
\renewcommand{\thetable}{S\arabic{table}}
\setcounter{table}{0}
\makeatletter
\c@secnumdepth = 2
\makeatother

\onecolumngrid

\begin{center}
 {\large \textmd{Supplemental Materials:} \\[0.3em]
 {\bfseries Even-odd parity transition in strongly correlated locally noncentrosymmetric superconductors : An application to CeRh$_2$As$_2$}}
 
\end{center}

\setcounter{page}{1}

\section{Self-consistent condition for fluctuation exchange approximation}

The noninteracting Green functions for $U=0$ are expressed
by the $4\times4$ matrix form in the spin and sublattice basis,
\begin{equation}
  \label{num:noint_green}
  G^{(0)}(\bm{k},i\omega_n) = \left(i\omega_n s_0\otimes\sigma_0 - \mathcal{H}_0(\bk)\right)^{-1},
\end{equation}
where $\omega_n=(2n+1)\pi T$ are fermionic Matsubara frequencies.
In the interacting case $U\neq0$, the dressed Green functions contain a self-energy, $\Sigma(\bm{k},i\omega_n)$, 
\begin{align}
  \label{num:green_function_dressed}
  G(\bm{k},i\omega_n) & = \left(i\omega_n s_0\otimes\sigma_0 - \mathcal{H}_0(\bk) -\Sigma(\bm{k},i\omega_n)\right)^{-1}.
\end{align}
Within the FLEX approximation, the self-energy is expressed with use of an effective interaction, $\Gamma^n(\bm{k},i\nu_n)$, as
\begin{align}
  \label{num:self_energy}
  &\Sigma_{\xi\xi'}  (\bm{k},i\omega_n) 
                          = \frac{T}{N} \sum_{\bm{q},i\nu_n}
  \Gamma^n_{\xi\xi_1\xi'\xi_2}(\bm{q},i\nu_n)G_{\xi_1\xi_2}(\bm{k}-\bm{q},i\omega_n-i\nu_n),
\end{align}
and the effective interaction is given by
\begin{equation}
  \label{num:effective_interaction}
  \Gamma^n_{\xi_1\xi_2\xi_3\xi_4}(\bm{k},i\nu_n) = U_{\xi_1\xi_2\xi_5\xi_6}\left(\chi_{\xi_5\xi_6\xi_7\xi_8}(\bm{k},i\nu_n)-\frac{1}{2}\chi^{(0)}_{\xi_5\xi_6\xi_7\xi_8}(\bm{k},i\nu_n)\right)U_{\xi_7\xi_8\xi_3\xi_4},
\end{equation}
where $U_{\xi_1\xi_2\xi_3\xi_4}$ is bare interaction tensor which satisfies the following relation
\begin{align}
    \sum_{\xi_1\xi_2\xi_3\xi_4} U_{\xi_1\xi_2\xi_3\xi_4} c^\dagger_{\xi_1} c_{\xi_2} c_{\xi_3} c^\dagger_{\xi_4} = U \sum_{i,\sigma} n_{i\uparrow \sigma}n_{i\downarrow \sigma}, \\
  U_{\xi_1\xi_2\xi_3\xi_4} = \delta_{\sigma_1,\sigma_2}\delta_{\sigma_2,\sigma_3}\delta_{\sigma_3,\sigma_4} U_{s_1 s_2 s_3 s_4}, \\
  U_{\uparrow\downarrow\uparrow\downarrow} = U_{\downarrow\uparrow\downarrow\uparrow} = -U_{\uparrow\uparrow\downarrow\downarrow} = -U_{\downarrow\downarrow\uparrow\uparrow} = U,
\end{align}
and $i\nu_n$ are bosonic Matsubara frequencies.
Here, $\chi(\bm{k},i\nu_n)$ is the generalized susceptibility.
We introduce the bare susceptibility
\begin{align}
  \label{num:bare_suscep}
  \chi^{(0)} _{\xi_1\xi_2\xi_3\xi_4}(\bm{q},i\nu_n)= -\frac{T}{N}\sum_{\bm{k},i\omega_n}G_{\xi_1\xi_3}(\bm{k},i\omega_n) G_{\xi_4\xi_2}(\bm{k}-\bm{q},i\omega_n-i\nu_n),
\end{align}
and compute the generalized susceptibility by
\begin{align}
  \label{num:gener_suscep}
  \chi_{\xi_1\xi_2\xi_3\xi_4}(\bm{q},i\nu_n)=\chi^{(0)}_{\xi_1\xi_2\xi_3\xi_4}(\bm{q},i\nu_n)+\chi^{(0)}_{\xi_1\xi_2\xi_5\xi_6}(\bm{q},i\nu_n)U_{\xi_5\xi_6\xi_7\xi_8}\chi_{\xi_7\xi_8\xi_3\xi_4}(\bm{q},i\nu_n).
\end{align}
According to Eqs.~(\ref{num:green_function_dressed})-(\ref{num:gener_suscep}),
$G$, $\Sigma$, $\Gamma^n$, $\chi^{(0)}$, and $\chi$
depend on each other, and therefore, we self-consistently determine these functions.

For functions with fermionic Matsubara frequencies $A(\bm{q},i \omega_n)$, the static limit 
$A(\bm{q},0)$ is evaluated by an approximation justified at low temperatures,
\begin{equation}
  A(\bm{q},0) \simeq \frac{A(\bm{q},i\pi T)+A(\bm{q},-i\pi T)}{2}.
\end{equation}

\section{linearized \'{E}liashberg equation}

To investigate superconductivity, we numerically solve
the linearized \'{E}liashberg equation which is given by
\begin{align}
  \lambda\Delta_{\xi\xi'}(k)  =
  \frac{T}{N}\sum_{k'} \Gamma^a_{\xi \xi_1 \xi_2 \xi'}(k-k')G_{\xi_1 \xi_3}(k')\Delta_{\xi_3 \xi_4}(k') G_{\xi_2 \xi_4}(-k'),
\end{align}
where $\Delta$ is the gap function and $\Gamma^a$ is
obtained by
\begin{align}
  \Gamma^a_{\xi_1\xi_2\xi_3\xi_4}(k-k') & = U_{\xi_1\xi_2\xi_3\xi_4}/2 +U_{\xi_1\xi_2\xi_5\xi_6}\chi_{\xi_5\xi_6\xi_7\xi_8}(k-k')U_{\xi_7\xi_8\xi_3\xi_4}.
\end{align}
Here we adopted abbreviated notation $k=(\bm{k},i\omega_n)$. 
Evaluating $\lambda$, eigenvalues of
the linearized \'{E}liashberg equation, we determine the
critical temperature $T_c$ from the criterion $\lambda=1$.

\section{Multipole susceptibility}

\begin{figure}[tbp]
 \begin{center}
    \includegraphics[keepaspectratio, scale=0.7]{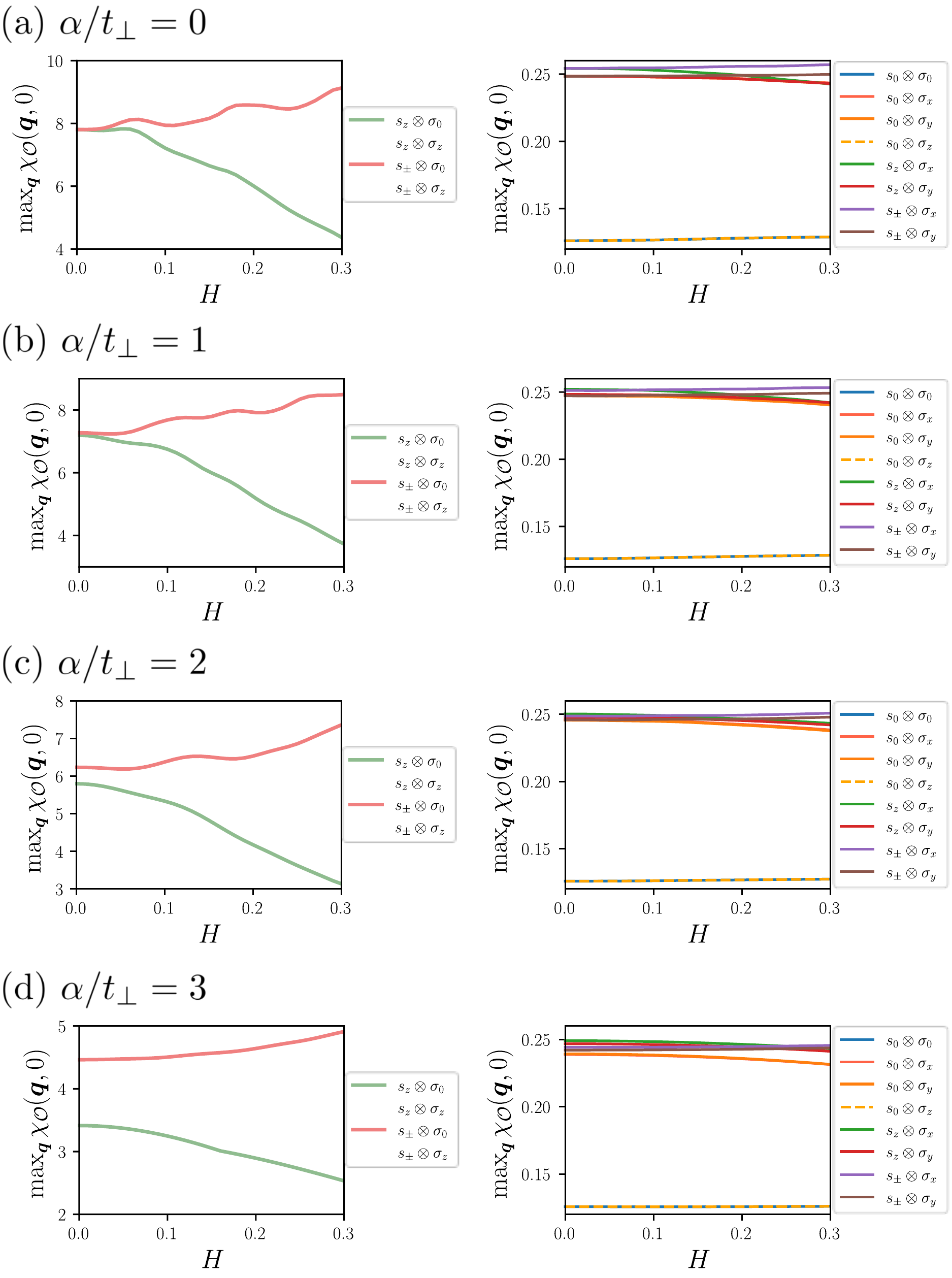}
  \end{center}
  \caption{ 
  (a-d) The magnetic field dependence of static multipole fluctuations for $\alpha/t_{\perp}=0,1,2,3$.
  We set $T=0.01$.
  }
  \label{fig:suppl_suscep}
\end{figure}

In Fig.~\ref{fig:suppl_suscep}, we show the magnetic field dependence of the maximum of multipole susceptibility for $\alpha/t_{\perp}=0,1,2,3$.
Figure~\ref{fig:suppl_suscep}(c) is the same as Fig.~2(a) in the main text.
In the left panels, transverse and longitudinal magnetic susceptibilities are shown.
In the right panels, the other multipole susceptibilities are shown.
Regardless of the value of $\alpha/t_{\perp}$, the transverse (longitudinal) magnetic susceptibility is enhanced (reduced) by the magnetic field, and the susceptibilities for other multipole operators are negligibly small.

\section{Eigenvalues of \'{E}liashberg equation}

\begin{figure}[tbp]
 \begin{center}
    \includegraphics[keepaspectratio, scale=0.7]{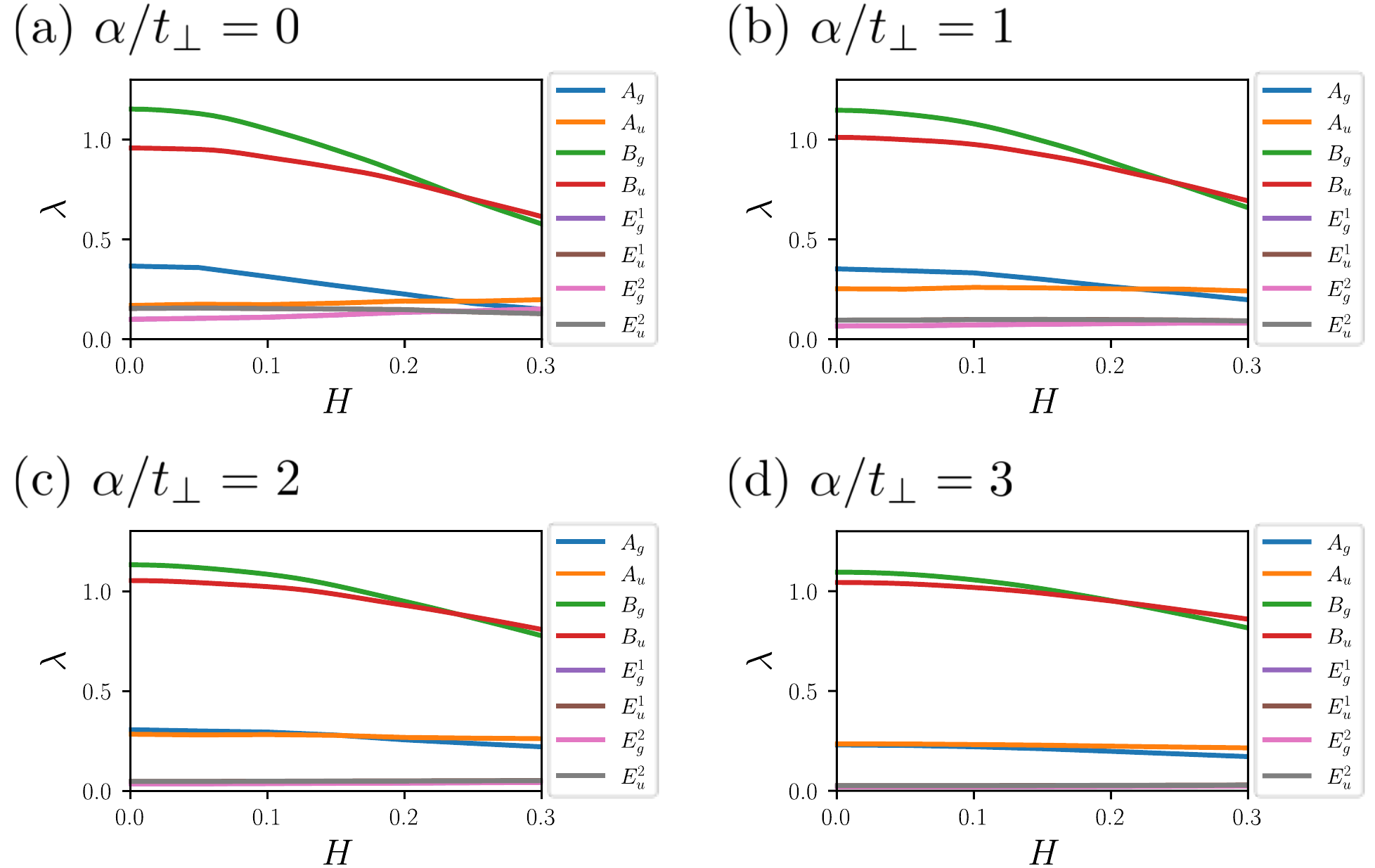}
  \end{center}
  \caption{ 
  (a-d) The magnetic field dependence of the eigenvalues of \'{E}liashberg equation for each irreducible representation. We adopt $\alpha/t_{\perp}=0,1,2,3$ and fix the temperature $T=0.01$.
  }
  \label{fig:suppl_lambda}
\end{figure}

In Fig.~\ref{fig:suppl_lambda}, we show the magnetic field dependence of the eigenvalues of \'{E}liashberg equation for $\alpha/t_{\perp}=0,1,2,3$.
Figure~\ref{fig:suppl_lambda}(c) is equivalent to Fig.~3(a) in the main text.

\section{\'{E}liashberg equation without field-enhanced magnetic anisotropy}

\begin{figure}[tbp]
 \begin{center}
    \includegraphics[keepaspectratio, scale=0.7]{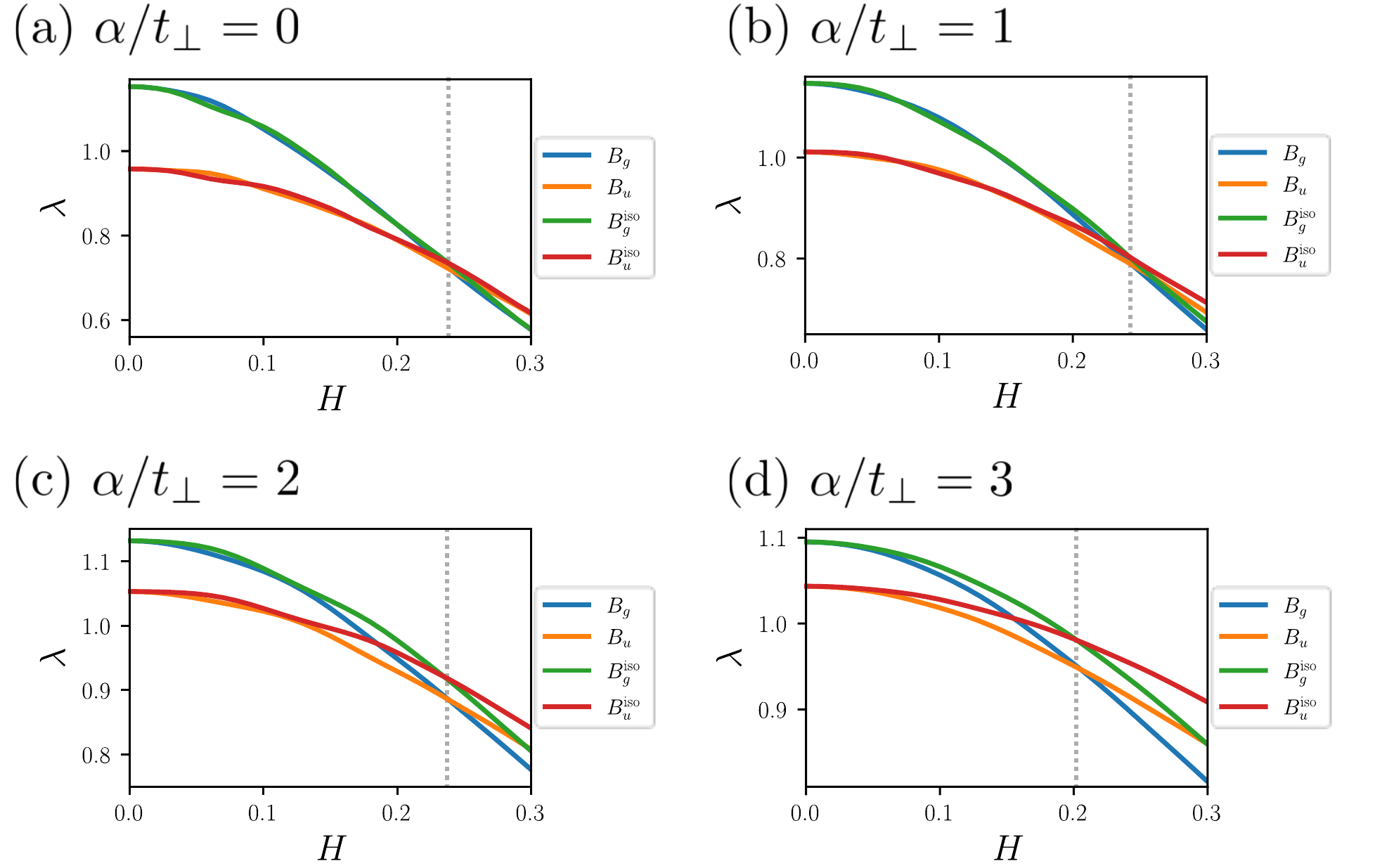}
  \end{center}
  \caption{ 
  (a)-(d) The magnetic field dependence of the eigenvalues of \'{E}liashberg equation for the $B_g$ and $B_u$ representations. We fix the temperature $T=0.01$ and show the results for $\alpha/t_{\perp}=0,1,2,3$. 
  $B^{\mathrm{iso}}_g$ and $B^{\mathrm{iso}}_u$ represent the eigenvalues obtained by neglecting the field-enhanced magnetic anisotropy. Details of the calculations are described in the main text.
  The parity transition points are indicated by the gray dashed lines.
  }
  \label{fig:suppl_iso}
\end{figure}

In Fig.~\ref{fig:suppl_iso}, we show the magnetic field dependence of the eigenvalues of \'{E}liashberg equation with and without field-enhanced magnetic anisotropy for $\alpha/t_{\perp}=0,1,2,3$.
Figure~\ref{fig:suppl_iso}(c) is shown as Fig.~6 in the main text.
As shown by the gray dashed lines, the parity transition points are not affected by the field-enhanced magnetic anisotropy.
We see that the effect of 
the field-enhanced magnetic anisotropy becomes remarkable with increasing the spin-orbit coupling $\alpha$. Thus, the field dependence of the pairing interaction is closely related to the spin-orbit coupling.

\end{document}